\documentclass{ws-ijmpa}
\newcommand{\be}{\begin{eqnarray}}
\newcommand{\ee}{\end{eqnarray}}
\newcommand{\la}{\langle}
\newcommand{\ra}{\rangle}

\begin{document}


%
%
\title{INSTANTON-INDUCED CORRELATIONS IN HADRONS}

\author{\footnotesize PIETRO FACCIOLI\footnote{email: faccioli@ect.it}}

\address{ E.C.T.$^\star$, 286 Strada delle Tabarelle, Villazzano
 (Trento) Italy, I-38050 and\\
 INFN, Gruppo Collegato di Trento.}

\maketitle


\begin{abstract}
QCD instantons generate non-perturbative 
spin- and flavor- dependent forces between quarks.
We review the results of a series of studies 
on the role played by instanton-induced correlations in hadrons. 
We first present a study of instanton-mediated interactions in
QCD, based on lattice simulations.
Then we show that the Instanton Liquid Model can reproduce  
the available data on proton and pion form factors at large momentum transfer.
The virial expansion in the vacuum diluteness parameter can explain why
the perturbative regime sets in very early in some physical processes and much
later in some other.
We discuss the connection with diquarks. Instantons generate a deeply bound 
scalar color anti-triplet diquark, with a mass of about 450~MeV and electric
charge radius~$\sim~0.7$~fm.
The strong attraction in the ${\bf \bar{3}_c}$ $0^+$ 
diquark channel leads to a quantitative
description of non-leptonic decays of hyperons and provides a microscopic
dynamical explanation of the $\Delta~I=1/2$~rule.
\end{abstract}

\section{Non-perturbative correlations in hadrons}

Several recent experiments have 
shown that strong non-perturbative forces inside 
hadrons survive also at relatively large momentum 
transfer, $Q^2\gtrsim~2-6~\textrm{GeV}^2$. 
Evidence for this fact has come, for example, from 
the measurement of the pion space-like form factor up 
to $Q^2\simeq~2~\textrm{GeV}^2$~\cite{jlabpion}, 
and from 
the direct determination of the proton $G_E(Q^2)/G_M(Q^2)$ ratio up to 
$Q^2\simeq~6~\textrm{GeV}^2$~\cite{JLABGEGM}.
In these reactions, the data 
are very far from the asymptotic perturbative QCD prediction. 
This behavior strongly contrasts with the observed perturbative scaling
of the DIS structure functions, for $Q^2\gtrsim~1~\textrm{GeV}^2$. 

It is tempting to argue 
that the delay in the onset of the perturbative regime
in the electro-magnetic form factors simply reflects the fact that
exclusive processes are very sensitive to the 
non-perturbative dynamics of hadronization.
However, this explanation is ruled-out by experiment: 
in fact, we know of at least one exclusive observable, 
the $\gamma\,\gamma^*\to~\pi_0$ transition
form factor, which gently follows the perturbative prediction, already 
for $Q^2\gtrsim~1-2~\textrm{GeV}^2$.

The JLAB results on form factors have provided new important insight
on the interactions inside hadrons. 
The evidence for
non-perturbative correlations at rather short distances, 
of the order of the tenth of the fermi, rules-out
models in which the quarks 
move essentially as free particle inside a confining bag and feel the 
non-perturbative color field only when
they approach the edge of the hadron, i.e. for distances of the order
$\sim~1$~fm.

From the comparison of the pion space-like form factor 
(which deviates from the perturbative prediction) and the 
$\gamma\,\gamma^*\to \pi_0$ transition form factor (which follows the
perturbative prediction) we have to conclude that the
non-perturbative short-range interactions 
are {\it strongly channel dependent}: they are effective in some 
process, but they are almost absent in some other.
This property of the non-perturbative dynamics has been known for a long
time: another  example is the "Zweig rule"~forbidding 
flavor-mixing, which works very well in the vector 
and axial vector meson channels, but it is violated in the scalar and 
pseudo-scalar channels.

The channel-dependence of the non-perturbative correlations 
implies that the quark-quark 
interaction at low-energy 
has much more structure than a simple radial potential. 
It must at least be spin-dependent. Moreover, the rather large
flavor asymmetry observed in  DIS parton distributions 
suggests that it is also  flavor-dependent.
As a consequence,  quarks are more correlated in some spin and flavor 
configuration than in other. Indeed, a number of
phenomenological studies seem to indicate that quarks preferably correlate
to form color- and flavor- anti-triplet scalar diquarks~(see \cite{exotica}
and references therein).

An important testing ground for the spin-flavor structure of the
non-perturbative quark-quark interaction is represented by
weak-decays of hadrons.
The natural scale of weak processes 
$-$set by  $W$ boson mass$-$  is much larger than all other 
scales involved in the hadron internal dynamics. 
Hence, weak interactions can be regarded as effectively local  and can 
 resolve small structures inside hadrons.
Moreover, their explicit dependence on flavor and chirality 
can be exploited
to probe the spin and flavor structure of the non-perturbative
QCD interaction.
Among the large variety of weak hadronic processes, a prominent role
is played by the 
non-leptonic decays of kaons and hyperons, which are characterized
by the famous $\Delta~I=1/2$ rule.
Neither electro-weak nor {\it perturbative} QCD interactions can account for 
the dramatic relative enhancement of the $\Delta~I=~1/2$ decay channels. 
Its origin must therefore reside in the non-perturbative sector of QCD.
Stech, Neubert and collaborators observed that the experimental data on  
both hyperon and kaon decays could be understood
by assuming strong diquark correlations
in the scalar, color anti-triplet channel~\cite{stech}.

The evidence for short-range, spin-dependent non-perturbative
correlations in hadrons naturally leads to the problem of identifying their
dynamical origin.
To this end, it is instructive to analyze the non-perturbative
scales in QCD.
We know of at least two non-perturbative
phenomena which occur at a momentum scale higher than $\Lambda_{QCD}$: 
the dynamical breaking of chiral symmetry and the anomalous 
breaking of the axial symmetry.
The natural scale for the interactions related to chiral symmetry breaking 
is set by the mass of the lightest vacuum excitation which is not protected by
chiral symmetry, the $\rho$-meson. Similarly, the typical
scale of topological interactions is given by the mass of the $\eta'$ meson. 
Hence, we 
should not be surprised to find  non-perturbative effects at the GeV scale.

The physical properties of the pion are certainly strongly influenced by
the interactions responsible for the breaking of chiral symmetry. 
Presumably, these forces play  an important role in the lightest baryons.

\section{Why instantons?}

Instantons have been proposed long ago as the dynamical mechanism
driving both the saturation of the axial anomaly~\cite{'thooft} and 
the spontaneous breaking of chiral 
symmetry~\cite{shuryak82}\,\cite{diakonov86} (~for a review
see~\cite{shuryakrev}~). The this hypothesis has been 
checked in a  number of lattice-based studies~\cite{lattice}.

Physically, instantons are  gluon fields which are
generated during tunneling events between degenerate 
QCD vacua.
Mathematically, they are non-perturbative solutions
of the Euclidean Yang-Mills equation of motion.
Being minima of the Yang-Mills action, they have been used
in the context of a saddle-point (semi-classical) 
analysis of the Euclidean QCD path 
integral~\cite{'thooft}.

In the Instanton Liquid Model (ILM)  the functional integral over all possible
gluon field configurations is replaced by a sum over the configurations of an 
ensemble of instantons and anti-instantons.
The path integral can then be solved by exploiting the formal 
analogy between the Euclidean 
QCD generating functional and the partition function
of a grand-canonical statistical ensemble.
The phenomenological parameters in the model are the average instanton
density $\bar{n}\simeq~1~\textrm{fm}^{-4}$$-$~which relates to 
the rate of tunneling in the vacuum~$-$
and the average instanton
size  $\bar{\rho}\simeq~1/3$~fm~$-$~which determines how long 
each tunneling event lasts for~$-$ . 
These values were extracted  more than two decades ago from the global 
vacuum properties~\cite{shuryak82} and indicate that
the diluteness (or ``packing fraction'') of the instanton ensemble 
is a small parameter: $\kappa=\bar{n}\,\bar{\rho}^4\simeq~0.01$.
Tunnelings are fairly rare events.

The instanton  field  induces an effective 
vertex between quarks ('t~Hooft interaction) which,
for $Q^2\ll~1/\bar{\rho}^{2}$, reduces to  a $2~N_f$-leg contact 
interaction.  For example, for $N_f=2$ it reads:
\be
\label{Lthooft}
\mathcal{L}_{'t~H}~=~G_{\bar{\rho},\bar{n}}\,
\left([\,(\psi^\dagger\,\tau^-_a\,\psi)^2 
- (\psi^\dagger\,i\gamma_5\,\tau^-\,\psi)^2\, ]
+\frac{1}{2(2\,N_c-1)}\,(\psi^\dagger
\,\sigma_{\mu\,\nu}\,\tau_a^-\,\psi)^2\right),\nonumber\\
\ee 
where $\tau^{-}=(\vec{\tau},i)$ ($\vec{\tau}$  are isospin Pauli matrices), and
$G_{\bar{n}\,\bar{\rho}}$ is a coupling constant depending on
the typical density and size of instantons in the vacuum. The  
finite size of the instanton field provides a natural cut-off scale 
for the 't~Hooft interaction.

The non-perturbative instanton-induced interaction (\ref{Lthooft})
has two important characteristic features: (i) it involves
quarks of different flavor, (ii) it is chirality-mixing, i.e.
quarks must flip their chirality any time they cross the field 
of an instanton. 
These two properties distinguish the instanton-induced interaction from 
the perturbative quark-gluon vertex, which is flavor-blind and 
chirality-conserving.

The chirality-flipping nature of the 't~Hooft vertex allows
to understand why {\it instanton effects are strongly channel dependent.}
In fact,  processes in which a quark undergoes a chirality-flip can 
occur any time there is an instanton nearby. In this case, the 
corresponding matrix elements will get a contribution 
from the instanton-induced interaction
at the leading order in the instanton packing fraction, i.e.~$o(\kappa)$.
Conversely,  transitions in which quarks do not change their chirality
get contribution from instantons only at the next-to-leading order in 
the packing fraction, i.e. $o(\kappa^2)$.
The physical reason is that the probability for two  tunneling 
events to occur during the same scattering process is suppressed, if tunnelings
are  sufficiently rare events.

This mechanism explains the suppression of non-perturbative
effects in several channels. A good example is
the suppression of flavor-mixing in the
 vector and axial-vector meson channels (Zweig rule).
The Lagrangian (\ref{Lthooft}) induces flavor-mixing. 
Due to its chirality-flipping structure, its
contribution to the scalar and pseudo-scalar channels 
is of $o(\kappa)$, so maximally enhanced. 
On the other hand, its contribution to the
vector and axial-vector channels is of
$o(\kappa^2)$, so much smaller.
 
The chirality-flipping structure of the 't~Hooft Lagrangian (\ref{Lthooft})
has been used to look for signatures of instanton-induced interaction
by means of lattice QCD simulations~\cite{scalar}.
If the  interaction between light quarks is mainly
instanton-mediated, then we expect a new chirality-flip any time 
quarks cross the field of an instanton.
Based on this observation, one is led to consider 
the ``chirality-flip'' asymmetry constructed by taking the ratio 
of the amplitude for a 
$|u\,\bar{d}\ra$ pair to be found after an interval $\tau$ in a state
in which the chiralities of the quark and the anti-quark 
are interchanged, relative to the amplitude
to remain in the same chirality state~\cite{chimix}.
It can be shown that, in the quantum field theory formalism, 
such a ratio reads:  
\be
\label{RNS}
R(\tau)=\frac{A_{flip}(\tau)}{A_{non-flip}(\tau)}=
\frac{\Pi_{PS}(\tau)-\Pi_S(\tau)}
{\Pi_{PS}(\tau)+\Pi_S(\tau)},
\ee
where 
\be
\Pi_{PS}(\tau)&=&\langle 0|J_{PS}(\tau)J^\dagger_{PS}(0) |0 \rangle,
\qquad
J_{PS}(x)=\bar{u}(x) \,i\gamma_5\,d(x),\\
\Pi_{S}(\tau)&=&\langle 0|J_S(\tau)J^\dagger_S(0) |0 \rangle,
\qquad
J_S(x)=\bar{u}(x)\,d(x).
\ee
Notice that the ratio $R(\tau)$ 
must vanish as $\tau\to 0$ (no chirality flips), and must approach 1 as  
$\tau\to \infty$ (infinitely many chirality flips). 

The instanton picture gives a very specific prediction for the 
asymmetry~(\ref{RNS}). If quarks propagate in the vacuum for a time 
comparable with the typical interval  between 
two consecutive tunneling events,  
they have a large probability of crossing the field of an instanton.
Hence, after some time, the quarks are most likely to be found in the
configuration in which their chirality is flipped and $R(\tau)>1$.
However, if one waits longer, then the quarks will ``bump''
 into another instanton field, which will re-flip their chirality and 
restore the initial chirality state.
So, the probability to find the quark and antiquark in the flipped chirality
state will start decreasing.
Hence, the instanton picture predicts that
the function $R(\tau)$ will have a maximum at $\tau~\sim\bar{n}^{1/4}$.

\begin{figure}
\centerline{\psfig{file=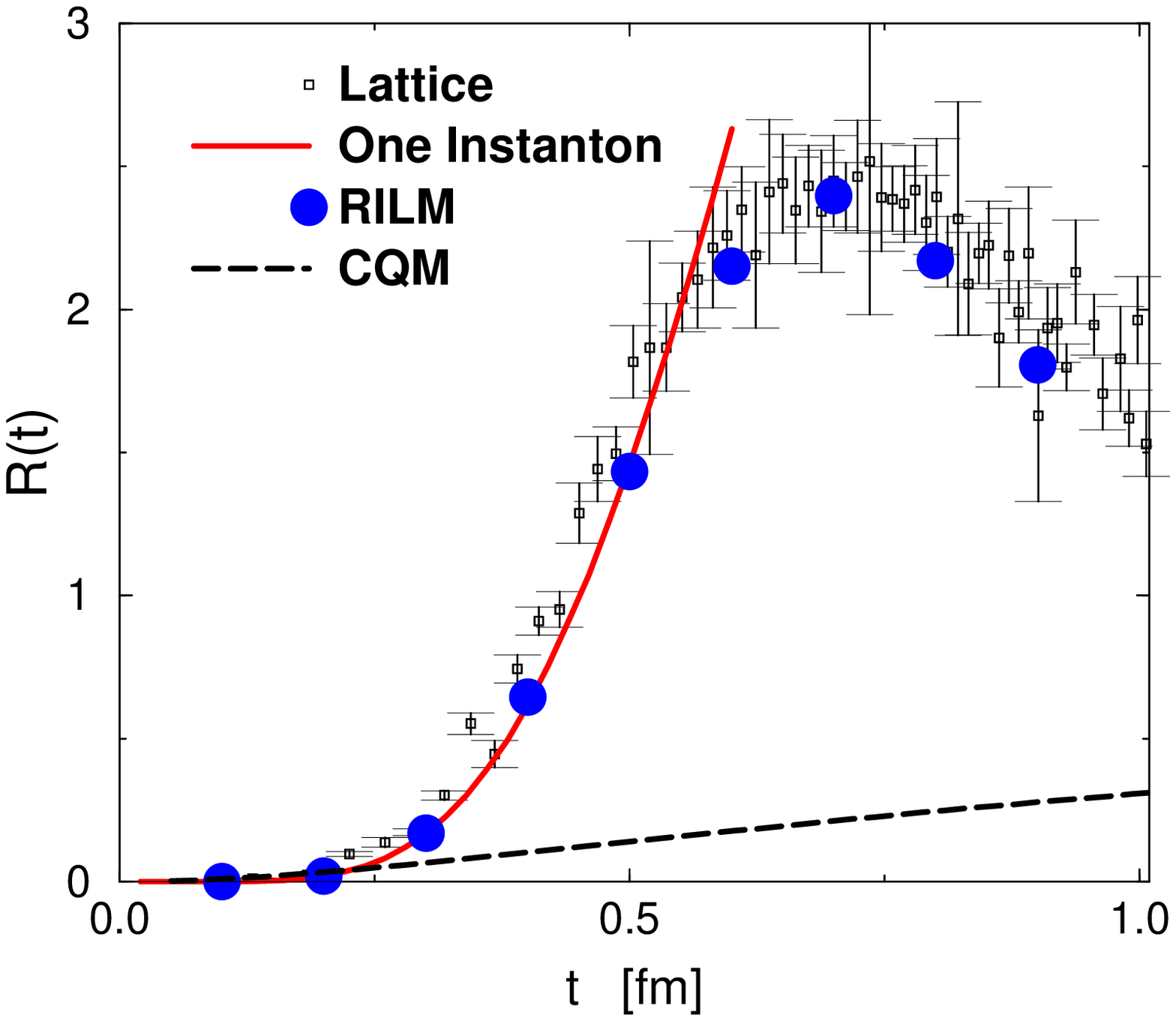,width=6cm,clip=}
\hspace{1.3cm}\psfig{file=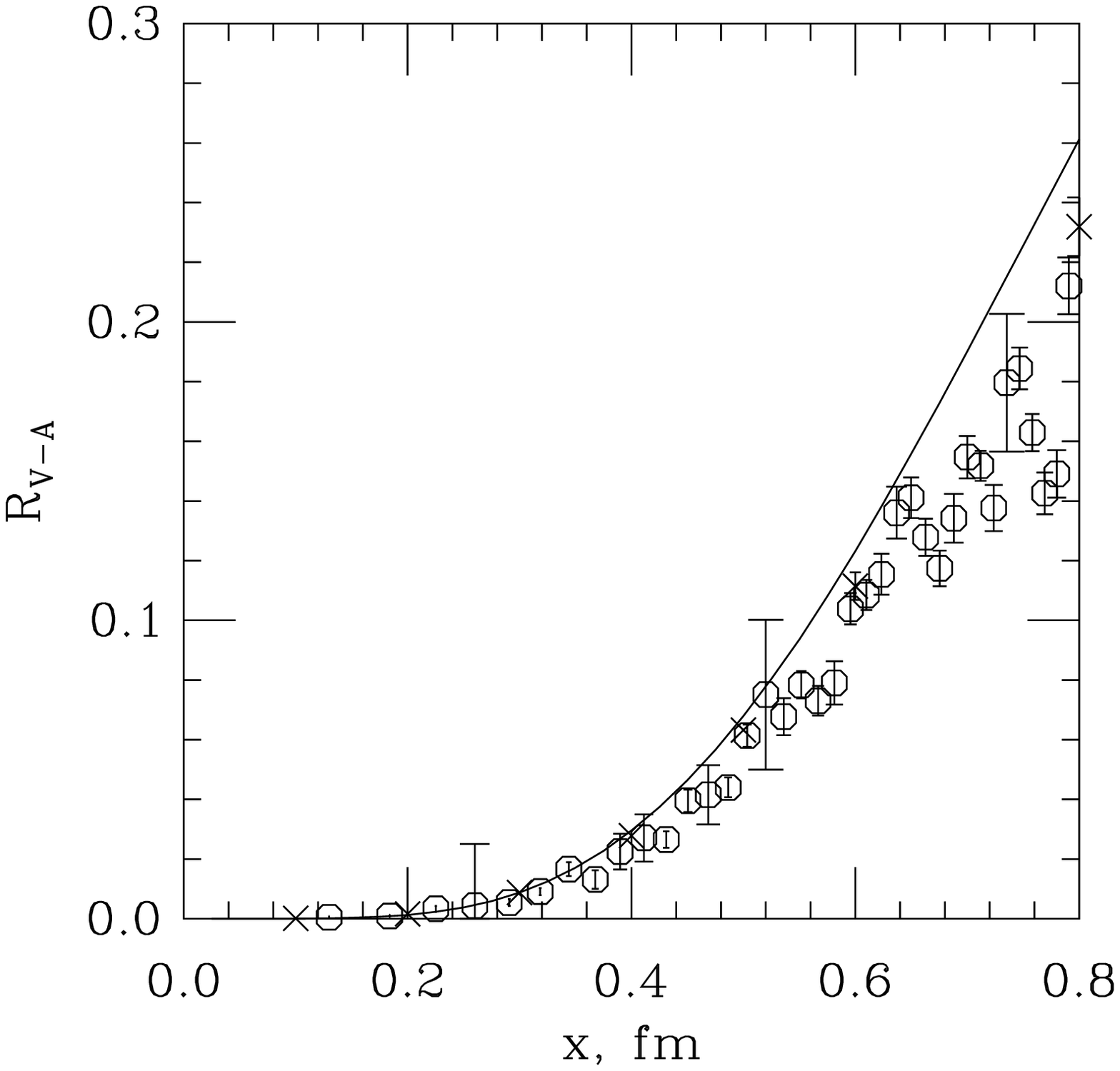,width=5.6cm,clip=}}
\caption{\mbox{}\protect\\
LEFT PANEL: The chirality-flip correlator (\ref{RNS})
in lattice QCD (square points) and in the ILM (circles).
The solid line represent the single-instanton contribution. 
The dashed curve was obtained from  two free ``constituent''
quarks with a mass of
400 MeV.\protect\\
RIGHT PANEL: The ratio (\ref{RAV}), evaluated 
from lattice QCD (octagons) -extrapolated to zero quark mass-,  
from the ILM (crosses) and  extracted from the ALEPH $\tau-$lepton decay data, 
as extracted in Ref.~\protect\cite{ALEPH}(lines).}
\label{chifig}
\end{figure}

Such a characteristic prediction of the instanton model has been checked
on the  lattice.
In Fig.~\ref{chifig}~(left panel) we compare the results of a quenched 
ILM simulation with the result of a quenched lattice 
calculation, performed using the same bare quark masses and the same 
four-dimensional box.
The agreement between the ILM and QCD is impressive, even at the quantitative 
level  (note that no-parameter fitting was involved in the ILM calculation).

A similar quantitative agreement between ILM and lattice 
simulations has been observed by DeGrand also in the
asymmetry constructed with vector 
and axial-vector two-point correlation functions~\cite{av}:
\be
\label{RAV}
R(\tau)=\frac{\Pi_V(\tau)-\Pi_{AV}(\tau)}
{\Pi_{V}(\tau)+\Pi_{AV}(\tau)},
\ee
where 
\be
\Pi_V(\tau)&=&\langle 0|J_\mu(\tau)J^\dagger_\mu(0) |0 \rangle,
\qquad
J_\mu(x)=\bar{u}(x) \,\gamma_\mu\,d(x),\\
\Pi_{AV}(\tau)&=&\langle 0|J_{\gamma_{\mu 5}}
(\tau)J^\dagger_{\gamma_{\mu 5}}(0) |0 \rangle,
\qquad
J_{\gamma_{\mu 5}}(x)=\bar{u}(x)\,\gamma_\mu \gamma_5\,d(x).
\ee
The results are presented in Fig.~\ref{chifig}~(right panel). We recall
that the vector and axial-vector 
point-to-point correlators receive contribution from instantons only 
at $o(\kappa^2)$. 
It is quite remarkable that the instanton-induced effects provide 
the correct amount of non-perturbative correlations 
even when they  are relatively 
suppressed: there seems to be little
room for additional non-perturbative dynamics, in these
correlators.

This collection of lattice results represents a compelling evidence 
for large instanton contributions to the dynamics of {\it light quark}.
On the other hand, it should be mentioned that 
instantons couple much more weakly to {\it heavy quarks}, 
and that the discussion of their role in pure gluon-dynamics is 
still rather controversial.

From this discussion it follows that instantons represent natural candidates 
for the microscopic dynamical mechanism underlying the
short-range non-perturbative correlations in light hadrons.
In the following sections we shall show that the ILM
quantitatively reproduces the JLAB data on electro-magnetic form factors 
and explains why the perturbative regime sets-in very early
in DIS and in the $\gamma^*\,\gamma\to~\pi_0$ transition form factor. 
Moreover, we will also show that instantons generate a scalar, color 
anti-triplet diquark bound-state of roughly 450~MeV mass.
Such strong instanton-induced diquark 
correlations lead to a quantitative understanding of non-leptonic 
weak decays of hyperons and to an explanation of the $\Delta~I=1/2$ rule.

\section{Instantons and the electro-magnetic structure of hadrons}
\label{FF}

Let us first discuss
the instanton contribution to the
electro-magnetic form factors.
The framework to compute momentum-dependent hadronic matrix elements, from
vanishing to large  momentum transfer, has been recently 
developed in a number of 
papers~\cite{i3ptILM}\,\cite{nucleonGE}\,\cite{mymasses}\,\cite{pionFF}\,\cite{nucleonFF}.
At large momentum transfer the calculations of hadronic matrix
elements in the ILM can even be carried-out analytically, by means of the 
Single Instanton Approximation~(SIA)~\cite{SIA}. This approach 
exploits the fact that small-sized correlation functions
are dominated by the interaction of the quarks with a single instanton.
At small or vanishing momentum transfer, many-instanton effects are important
and one has to rely on numerical Monte Carlo methods to compute
the path integral and extract the matrix elements from
appropriate ratio of correlation functions~\cite{nucleonFF}.
\begin{figure}
\centerline{\psfig{file=Q2FF.eps,width=5.4cm,clip=}
\hspace{2cm}\psfig{file=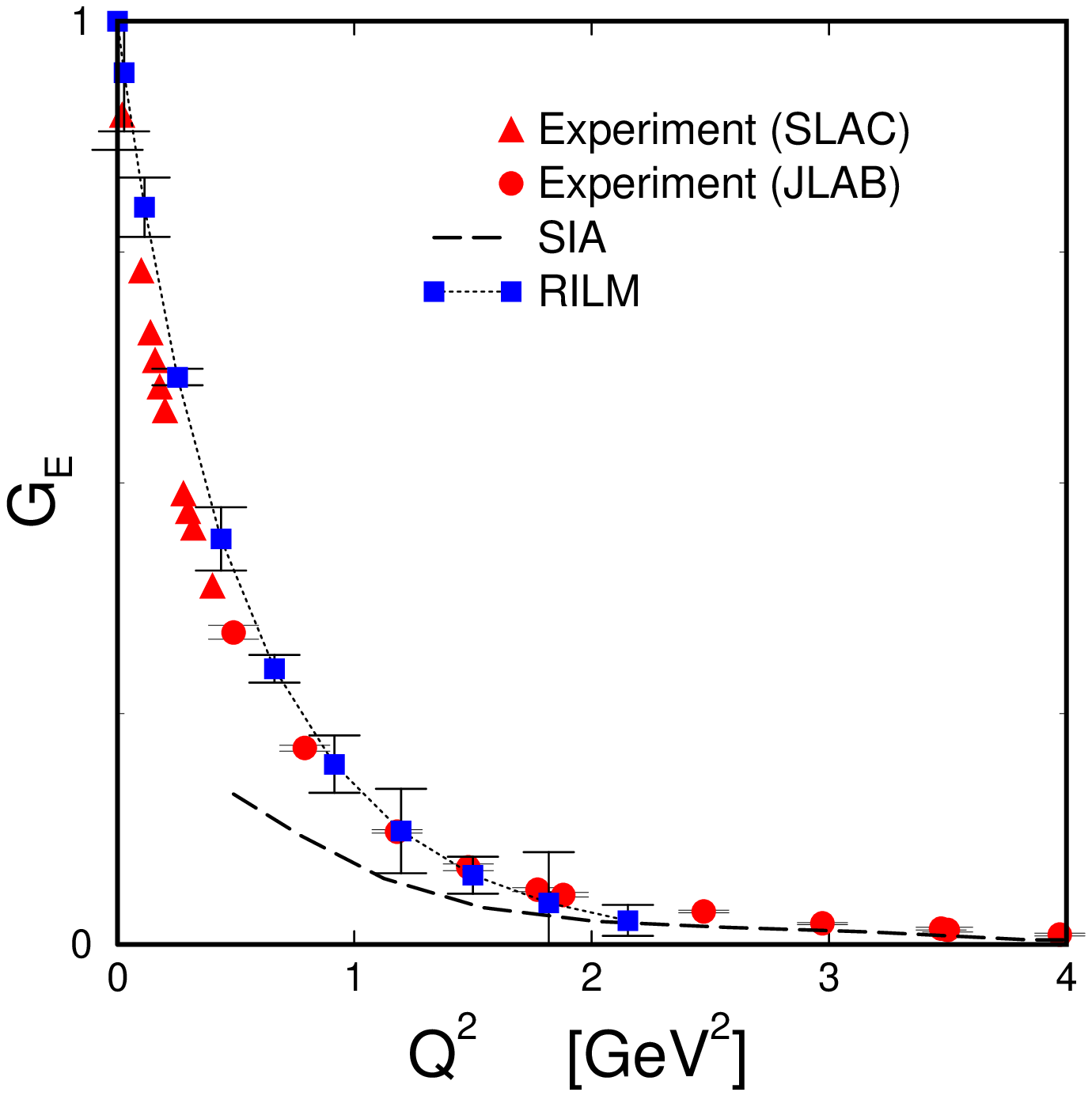,width=4.5cm,clip=}}
\caption{\mbox{}\protect\\
LEFT PANEL: The JLAB data for $Q^2 \, F_\pi(Q^2)$ in comparison with
the asymptotic perturbative QCD prediction (thick bar, for a typical $\alpha_s
\sim 0.2-0.4$), the monopole fit
(dashed line), and the SIA calculation (solid line). The SIA
calculation is not reliable below $Q^2 \sim 1 \, \textrm{GeV}^2$.
The solid circles denote the SLAC data.\protect\\
RIGHT PANEL: The electric form factor of the proton in the ILM and 
from experiment.
Triangles are low-energy SLAC data, which follow a dipole fit.
Circles are experimental data obtained from the
recent JLAB result for $G_E/G_M$, assuming a dipole fit for the 
magnetic form factor.
Squares are result of many-instanton simulations in the ILM, and the
dashed line is the SIA curve.}
\label{pionFFfig}
\end{figure}

The result of our analytic calculation of the pion form factor
at moderate and large momentum transfer~\cite{pionFF}
is reported in Fig.\ref{pionFFfig}. The instanton contribution
is $o(\kappa)$, i.e. maximally enhanced. The ILM
can quantitatively reproduce the available data and explain the deviation 
from the perturbative regime at large momentum transfer.  

Conversely,  we have observed that in the $\gamma\, \gamma^*\to \pi_0$  
transition form factor the instanton effects are of order
$o(\kappa^2)$, hence parametrically  
suppressed by an additional power of the packing fraction.
This explains the early onset of the perturbative regime in such a form factor.
Moreover, a calculation of the
pion light-cone distribution amplitude in the ILM 
was performed in \cite{Dorokhov}.
It was found that instantons can explain the behavior 
of the low-energy experimental data ($Q^2<~2$~GeV$^2$) 
for the  $\gamma\,\gamma^\star~\to~\pi^0$ transition form factor.

Our ILM calculation of the proton electric form factor $G_E(Q^2)$ is reported
in Fig.\ref{pionFFfig}~(right panel), where it is compared with a 
SLAC data at low $Q^2$ and of the JLAB data at large-$Q^2$.
Again, the short-range correlations induced by instantons quantitatively
explain the experimental data. Similar results have been obtained also for
 the magnetic form factor~\cite{nucleonFF}.

Physically,
the fact that instantons give very hard hadronic
form factors can be interpreted as 
follows. Due to the strong zero-mode attraction, 
the hadron wave-function in coordinate space
is very narrow and peaked around the origin. 
As a consequence, the charge distribution changes very rapidly at short 
distances~(see for example Fig.~\ref{diquarkfig}, right panel) leading to large
hard-momentum components in its Fourier transform (i.e. the form factor).
However, due to the finite size of the instanton, 
the zero-mode attraction cannot transfer infinitely large momenta. Hence,
the electric charge distribution has  to become eventually flat, for distances
much smaller than the instanton size.
Indeed,  when $Q^2\gtrsim~10-15~\textrm{GeV}^2$, we have observed that 
the zero-mode contribution to the electro-magnetic 
correlation function rapidly
dies-out and the perturbative regime is finally free to set-in. 
In conclusion, the ILM predicts that the perturbative limit
will not be reached,  in the kinematic region accessible to the 
forthcoming JLAB experiments.

Instantons also provide a possible explanation of why
DIS {\it leading-twist} perturbative evolution equations works so well, already for
$Q^2\gtrsim~1~\textrm{GeV}^2$.
Lee, Weiss and Goeke~\cite{hightwist}  have 
shown that the instanton-contribution
  to the twist 3 operators in the Operator Product Expansion
is parametrically suppressed by
powers of the packing fraction $\kappa$,
in complete
analogy with what we found in the $\gamma\,\gamma^*\to~\pi_0$ transition 
form factor.
Moreover, Ostrovsky and Shuryak have recently shown that the ILM
can explain  the available data on azimuthal spin asymmetries 
in DIS~\cite{ostro}.
\begin{table}
\tbl{dd}
{\begin{tabular}{@{}ccccc@{}} \toprule
Amplitude & P-wave (theory) & P-wave (experiment) 
&S-wave (theory) & S-wave (experiment) \\ 
(~$\times~10^{7}$~) \\\colrule
$\Lambda^0_0$&$-10.9\pm1.17$&$-15.61\pm1.4$&$-1.75\pm0.34$&$-2.36\pm0.03$\\
$\Lambda^0_-$&$17.71\pm1.66$&$22.40\pm0.54$&$2.25\pm0.57$&$3.25\pm0.02$\\
$\Sigma^+_0$&$22.4\pm3.55$&$26.74\pm1.32$& $-3.55\pm0.64$&$-3.25\pm0.02$\\
$\Sigma^+_+$&$31.84\pm4.81$&$41.83\pm0.17$& $0$        &$0.14\pm0.03$\\
$\Sigma^-_-$&$-1.52\pm0.30$&$-1.44\pm0.17$&$4.34\pm0.90$&$4..27\pm0.01$\\
$\Xi^-_-$&$14.15\pm2.75$&$17.45\pm0.58$&$-4.22\pm0.82$&$-4.49\pm0.02$\\
$\Xi^0_0$&$-10.42\pm1.95$&$-12.13\pm0.71$&$3.20\pm0.58$&$3.43\pm0.06$\\
\botrule
\caption{Random Instanton Liquid Model prediction and experimental results for
P-wave and S-wave amplitudes for non-leptonic weak decays of hyperons.
Following the standard notation, 
$B^Q_q$~corresponds to $\textrm{Amp}(B^Q\to B' + \pi^q)$. } 
\end{tabular}}
\label{resultdelta12}
\end{table}

\section{Instantons and the $\Delta{I}=1/2$ rule for hyperon decays}
\label{del}
As we have already mentioned, non-leptonic weak decays of hyperons are
good testing grounds for spin-dependent, non-perturbative
instanton correlations. 
The  decay 
amplitudes can be parametrized in terms of two constants, 
corresponding to parity-violating and parity-conserving transitions:
\be
\label{PSwave}
\la B'\,\pi| H_{eff} |B\ra = i\,\bar{u}_{B'}\,\left[A-B\gamma_5\right]\,u_B.
\ee
 $H_{eff}$ is the effective Hamiltonian, which incorporates the 
electro-weak and the hard-gluon contributions, 
$B$ ($B'$) denotes the initial  (final)
baryon, and $A$ and $B$ are respectively called
\emph{S-wave} and \emph{P-wave} amplitudes, each of which can be decomposed
in  $\Delta~I=1/2$ and $\Delta~I=3/2$ components. The $\Delta~I=1/2$
transition amplitudes are found to be typically $\sim~20$~times 
larger that the $\Delta~I=3/2$ amplitudes (``$\Delta~I=1/2$'' rule).

We have calculated these amplitudes in the ILM~\cite{delta12}. 
Our results are reported and compared with the experiment in 
table~\ref{resultdelta12}. We have found that not only instantons can explain
the $\Delta~I=1/2$ rule, but also that the theoretical predictions 
for the amplitudes were in quantitative agreement with experiment.
Note that a $\sim20\%$   discrepancy between theory and experiemt is of 
the same order of the uncertainty
which was introduced by assuming the flavor $SU(3)$ limit in the calculation.

In \cite{kochelev} Kochelev and Vento (KV) computed the instanton contribution 
to non-leptonic kaon decays.
On a qualitative level, they found a similar result: the 
inclusion of the instanton effects indeed produces
a strong enhancement of the $\Delta~I=1/2$ decay 
channel. On a quantitative level,
such an enhancement was found to be still insufficient 
to reproduce the experimental data.
However, it should be mentioned that non-leptonic
kaon decays in the $\Delta~I=1/2$ channel receive large contribution
also from final-state interactions, which have not been included in 
the KV analysis.
Moreover,  it is now clear that
the  KV calculation is undershooting the instanton 
contribution\footnote{The KV calculation 
was performed in the single-instanton approximation. 
In such an approach, one treats 
explicitly the degrees of freedom of the closest instanton and introduces an
additional parameter $m^*$, which effectively 
encodes contributions from all other instantons.
In their calculation, the authors adopted the phenomenological estimate
for $m^*$ which was available at the time, $m^*\simeq 260$~MeV.
Later, the same parameter was rigorously defined, 
and determined from numerical simulations in the ILM \cite{SIA}, 
It was found to be considerably smaller ($m^*\simeq~80$~MeV).}.

\section{Instantons and diquarks}
\label{Dq}

The instanton-induced interaction (\ref{Lthooft}) in particularly attractive
in  the $0^+$ anti-triplet diquark channel.
The question whether such interaction leads to binding has been first 
posed in~\cite{binding}. 
In an exploratory study based on numerical calculation of point-to-point
correlation functions~\cite{baryonRILM}, 
Shuryak, Sch\"afer and Verbaarshot found some indication of 
an instanton-induced deeply bound $0^+$ anti-triplet diquark, 
with a mass of roughly 450~MeV.
On the other hand, Diakonov and collaborators
have analyzed the same channel by
solving Schwinger-Dyson equations at the mean-field level (leading order in
$1/N_c$)~\cite{diak}. 
They found evidence for correlations, but no 
binding\footnote{Except for $N_c=2$, in which case
the diquark is a baryon and its 
mass is protected by Pauli-Gursey symmetry.}. The role of instanton-induced
diquark correlations in the nucleon has been also investigated in the context
of QCD sum-rules~\cite{kochelev1}.

\begin{figure}
\centerline{\psfig{file=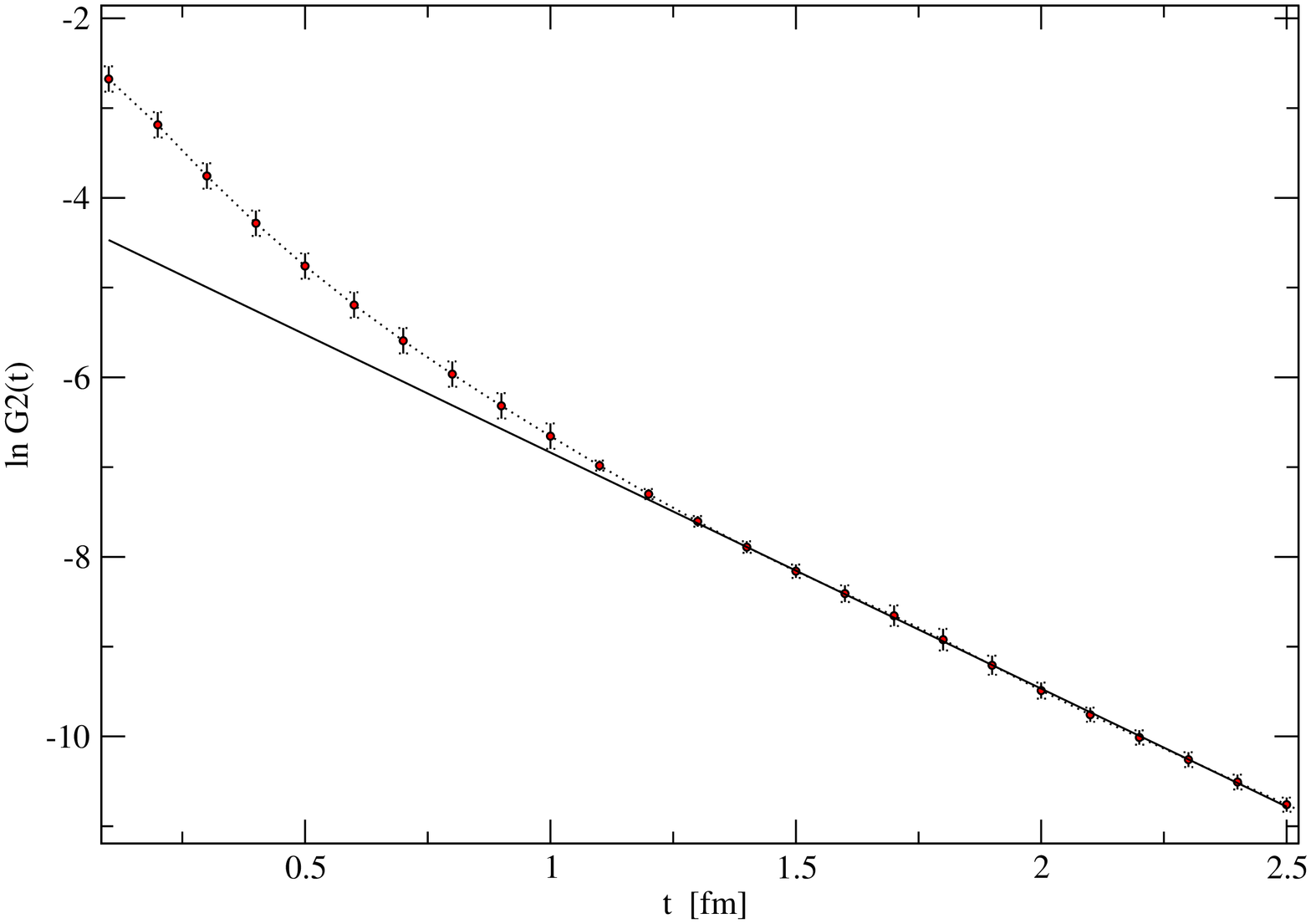,width=5.cm,height=6.3cm, angle=-90}
\hspace{0.8cm}\psfig{file=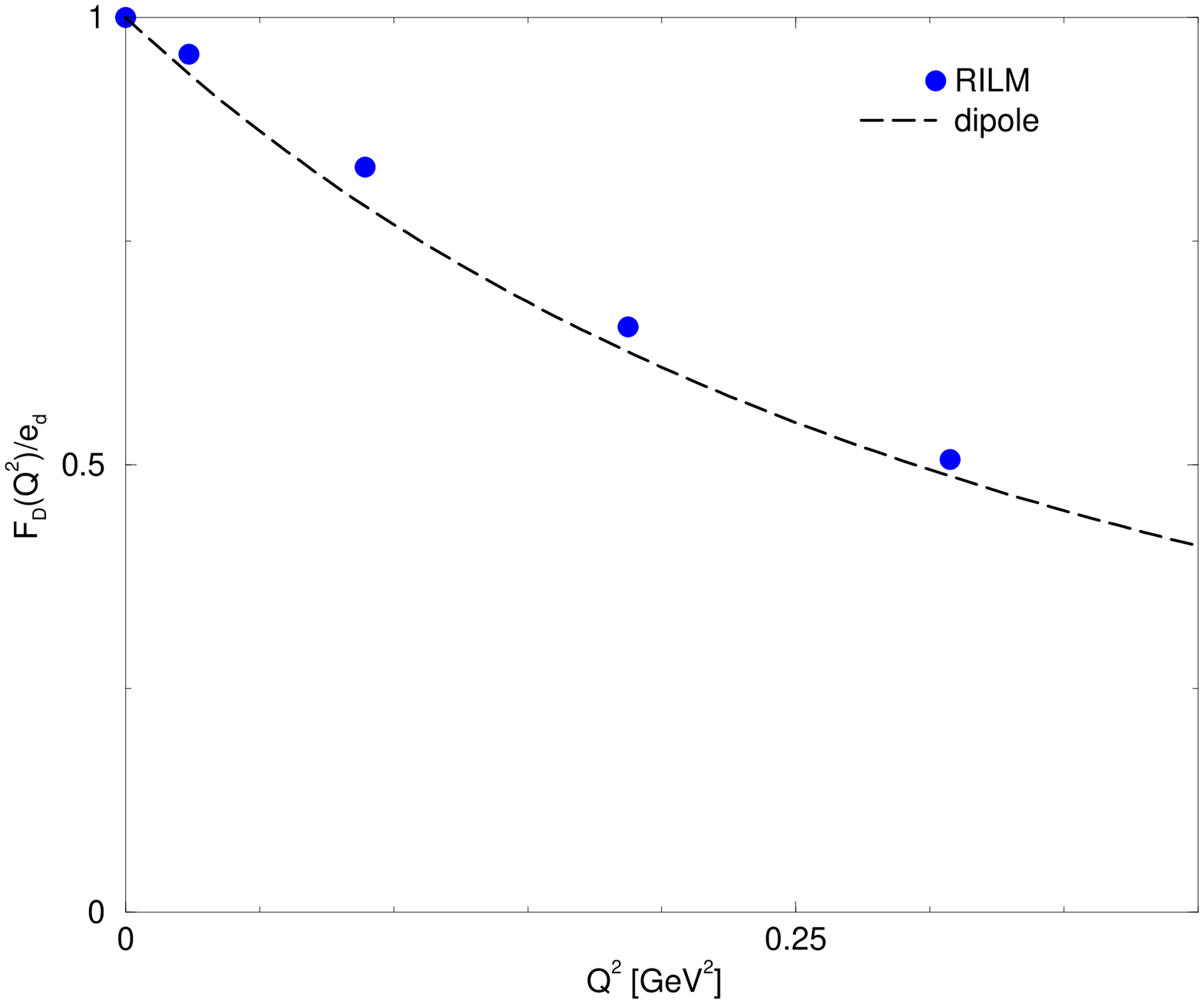,width=4.5cm,clip=, angle=-90}}
\caption{\mbox{}\protect\\
LEFT PANEL:
Logarithm of the diquark two-point function, $\ln G2(\tau)$ computed
in the Random Instanton Liquid Model. The linear slope is a clean 
signature of the existence of a bound state.\protect\\
RIGHT PANEL:
The diquark form factor (normalized to the total diquark charge) 
in the Random Instanton Liquid Model (points) 
compared with the dipole fit parametrizing low-energy data on 
the proton electric form factor,  $F_{dip}=1(1+Q^2/0.71)^2$ (dashed line).}
\label{diquarkfig}
\end{figure}

In order to clarify whether instantons do or do not generate a scalar
diquark bound-state, 
we have computed numerically the diquark 
correlation function 
\be
G2(\tau)=\int d^3 {\bf x} 
~\langle 0|~T[~J_D({\bf 0},\tau)J^\dagger_D({\bf x},0)\,
Pe^{\int dy_\mu A_\mu(y)}~]|0\rangle,
\ee
where $J_D(x)$ is the scalar 
color-antitriplet diquark interpolating operator and 
the path-ordered exponent has been inserted to assure gauge-invariance.
If there is a bound-state diquark, then  
in the large Euclidean time limit
the logarithm of the two-point function must scale linearly with $\tau$:
\be
\ln G2(\tau)\stackrel{\tau\to\infty}{\rightarrow} \alpha -\tau~M_D,
\ee
where $\alpha$ is a constant and $M_D$ should be smaller than twice
the constituent quark mass.
The result of our calculation (which accounts for 
the instanton-induced interaction to all orders and does not involve
the large $N_c$ limit) 
is presented in
 Fig.~\ref{diquarkfig}~(left panel).
Clearly, we found unambiguous evidence for a diquark bound-state 
with $M_D\simeq~450$~MeV~\cite{diquark}.

An important related problem is what is the size of such a diquark. To 
answer this question we have extracted the diquark electric 
charge radius from its
electro-magnetic form factor~\cite{diquark} 
(see Fig.~\ref{diquarkfig},~right panel). 
We found that $r_E\sim~0.7$~fm which implies that the size of the
diquark is comparable with that of the proton.
\section{Conclusions}

We have reported on the results of a series of investigations on 
instanton-induced correlations. 
We have discussed a collection of lattice results which provide 
indications for large instanton-induced contributions to 
the dynamics of {\it light quarks}.
We have shown that the ILM provides a qualitative and quantitative
understanding on the global electro-weak structure of the pion and the proton.
In particular, it can explain the fact that 
the short-range non-perturbative correlations are very strong in some 
reactions and almost absent in some others.

We have presented a calculation which clearly shows that 
instantons lead to the formation of a 
deeply-bound ($0^+$,  ${\bf \bar{3}_c}$)
diquark, with a mass of about 450 MeV and size of the same order 
of that of the proton.
The quantum numbers and the mass of the instanton-induced diquark 
make it a candidate for the ``good diquark'', claimed by Jaffe and 
Wilczek~\cite{exotica}.


\begin{thebibliography}{0}

\bibitem{jlabpion}
J. Volmer {\it et al.} [The Jefferson Laboratory $F_\pi$
Collaboration], Phys. Rev. Lett. \textbf{86} 
(2001) 1713.

\bibitem{JLABGEGM} M.K. Jones, \emph{et al.},
Phys. Rev. Lett. \textbf{84} (2000) 1398.
O. Gayou,  \emph{et al.}, Phys. Rev. Lett. \textbf{88} (2002) 092301.

\bibitem{exotica} R.L.~Jaffe, hep-ph/0409065

\bibitem{stech} B. Stech, Phys. Rev. D {\bf 36} (1987) 975.
B. Stech and Q.P. Xu,  Z. Phys. C {\bf 49} (1991) 491.
H.D. Dosch, M. Jamin and B. Stech, Z. Phys. C {\bf 42} (1989) 167.
M. Neubert, Z. Phys. C {\bf 50} (1991) 243.
M. Neubert and B. Stech, Phys. Rev. D {\bf 44} (1991) 775.

\bibitem{'thooft}
    G.'t Hooft, Phys. Rev. Lett., \textbf{37} (1976) 8.
    G.'t Hooft,    Phys. Rev.,  \textbf{D14} (1976) 3432.
    G.~'t Hooft, Phys.\ Rept.\  {\bf 142}, 357 (1986).


\bibitem{shuryak82} E.V. Shuryak,
Nucl. Phys. \textbf{B214} (1982) 237.

\bibitem{diakonov86}
  D.J. Dyakonov and V.Yu. Petrov, Nucl. Phys.,
   \textbf{B272} (1986) 457.

\bibitem{shuryakrev}
T. Sch\"afer and E.V. Shuryak,
Rev. Mod. Phys. \textbf{70} (1998) 323.


\bibitem{lattice}
T.~DeGrand and A.~Hasenfratz,
Phys.\ Rev.\ D {\bf 64}, 034512 (2001).
T.~DeGrand and A.~Hasenfratz,
Phys.\ Rev.\ D {\bf 65}, 014503 (2002).
I.~Hip, T.~Lippert, H.~Neff, K.~Schilling and W.~Schroers,
Phys.\ Rev.\ D {\bf 65}, 014506 (2002).
R.~G.~Edwards and U.~M.~Heller,
Phys.\ Rev.\ D {\bf 65}, 014505 (2002).
T.~Blum {\it et al.},
Phys.\ Rev.\ D {\bf 65}, 014504 (2002).
C.~Gattringer, M.~Gockeler, P.~E.~Rakow, S.~Schaefer and A.~Sch{\"a}fer,
Nucl.\ Phys.\ B {\bf 617}, 101 (2001).
S.~J.~Dong {\it et al.},
Nucl.\ Phys.\ Proc.\ Suppl.\  {\bf 106}, 563 (2002).



\bibitem{scalar}
P.~Faccioli and T.A.~DeGrand, Phys. Rev. Lett. \textbf{91} 
(2003) 182001.  


\bibitem{chimix}
P.~Faccioli,
hep-ph/0211383.


\bibitem{av}
T.~DeGrand,
Phys.\ Rev.\ D {\bf 64} (2001) 094508. 


\bibitem{ALEPH}
T.~Schafer and E.~V.~Shuryak,
Phys.\ Rev.\ Lett.\  {\bf 86}(2001) 3973. 

\bibitem{jw}
R.~L.~Jaffe and F.~Wilczek,
Phys.\ Rev.\ Lett.\  {\bf 91}(2003) 232003. 


\bibitem{i3ptILM}
P.~Faccioli and E.~V.~Shuryak,
Phys.\ Rev.\ D {\bf 65} (2002) 076002.

\bibitem{nucleonGE}
P.~Faccioli, A.~Schwenk and E.~V.~Shuryak,
Phys.\ Lett.\ B {\bf 549} (2002) 93.

\bibitem{mymasses}
P.~Faccioli,
Phys.\ Rev.\ D {\bf 65}(2002) 094014. 

\bibitem{pionFF} P. Faccioli, A. Schwenk and E.V. Shuryak,
Phys. Rev. {\bf D67} (2003) 113009.

\bibitem{nucleonFF}
P.~Faccioli,
Phys.\ Rev.\ C {\bf 69} (2004) 065211.  

\bibitem{SIA} P. Faccioli and E.V. Shuryak,
Phys. Rev. \textbf{D64} (2001) 114020.


\bibitem{Dorokhov}
A.~E.~Dorokhov,
JETP Lett.\  {\bf 77} (2003) 63
[Pisma Zh.\ Eksp.\ Teor.\ Fiz.\  {\bf 77}, 68 (2003)].

\bibitem{hightwist}
N.~Y.~Lee, K.~Goeke and C.~Weiss,
Phys.\ Rev.\ D {\bf 65} (2002) 054008. 

\bibitem{ostro}
D.~Ostrovsky and E.~Shuryak,
hep-ph/0409253.


\bibitem{delta12}
M.~Cristoforetti, P.~Faccioli, E.~V.~Shuryak and M.~Traini,
Phys. Rev. {\bf D70}(2004) 054016.

\bibitem{kochelev} N.I.~Kochelev and V.~Vento, Phys.~Rev~.Lett.~{\bf 87} 
(2001) 111601. 

\bibitem{binding}
D.~Diakonov and V.~Petrov, "Diquarks in the instanton picture", in
the proceedings of the conference ``Quark Cluster Dynamics'', Bad Honnef,
29 June - 1 July 1992, Springer.

\bibitem{kochelev1}
A.E.~Dorokhov and N.I.~Kochelev, Z. Phys. {\bf C46} (1990),
 281.

\bibitem{baryonRILM} 
T.~Schaefer, E.V.~Shuryak and J.~Verbaarschot,
Nucl. Phys. {\bf B412} (1994) 143. 

\bibitem{diak} D.~Diakonov, H.~Forkel and 
M.Lutz, Phys. Lett. {\bf B373} (1996) 147. 
D. Diakonov and G.W. Carter, Phys. Rev. {\bf D60} (1999) 016004.

\bibitem{diquark}
M.~Cristoforetti, P.~Faccioli, G.~Ripka and M.~Traini,
hep-ph/0410304.


\end{thebibliography}
\end{document}